\documentstyle[revtex]{aps}
\begin{document}
\draft
\def\helium{$^4\mskip-0.4mu{\rm He}$}
\hspace{5 in} {\bf Preprint IFUM 444/FT}

\hspace{5.4 in} {\bf April 1992}
\vspace{5 mm}

\begin{title}
 Critical exponents of the three-dimensional classical plane rotator\\
  model on the sc lattice from a high temperature series analysis\\
 \end{title}
\author{P. Butera and M. Comi}
\begin{instit}
Istituto Nazionale di Fisica Nucleare\\
Dipartimento di Fisica, Universit\`a di Milano\\
Via Celoria 16, 20133 Milano, Italy
\end{instit}
\author{A. J. Guttmann}
\begin{instit}
Department of Mathematics, The University of Melbourne\\
Parkville, Victoria 3052, Australia
\end{instit}
\begin{abstract}
High temperature series
expansions of the spin-spin correlation
function for the plane rotator (or XY) model
on the sc	lattice are  extended
by three terms through order
$\beta^{17}$. Tables of the expansion coefficients
are reported for the correlation function
spherical moments of order $l=0,1,2$.
Our analysis of the
series  leads to  fairly
accurate estimates of the
critical parameters.

\end{abstract}
 \vspace{5 mm}
%\pacs{Ms number XX9999. PACS numbers: 05.50+q, 64.60.Cn, 75.10.Hk}

In three dimensions the two-component vector  model
is the simplest spin model in the universality class
of the
 superfluid $\lambda$ transition  of \helium,
and of the ferromagnetic transition
 of   magnets with an easy magnetization plane[\cite{vaks}].
No high temperature (HT) series studies of this
model have appeared in the last two decades
in spite of remarkable  experimental
measurements of the critical parameters in superfluid \helium\
and intense theoretical activity in
Renormalization Group calculations and
by direct Monte Carlo simulations.

In particular we should mention that the  critical index $\nu$,
 which describes the leading singularity of the superfluid fraction
in \helium\ near the superfluid transition temperature,
 has been measured with high precision
in a long series of experiments by G. Ahlers and his
collaborators[\cite{ahlers}].
As stressed by Ahlers, the superfluid fraction
 is the most accurately known singular parameter
at a critical point, and correspondingly $\nu$ is the
 most accurately known critical index.
The most recent experiments yield the value
$\nu = 0.6705 \pm 0.0006$.

Unfortunately the critical exponent $\gamma$
cannot be measured in liquid
\helium\ and, as far as magnetic systems
are concerned, no precise measurements
exist either for $\gamma $ or for $\nu$.
A  review of static critical properties of \helium\
 can be found in Ref.[\cite{hohe}] and a
general discussion of the interpretation of the
measurements on \helium\ in connection
with the problem of confluent singularities is given in
Ref.[\cite{confl}].

The Hamiltonian  of the three-dimensional
plane rotator (or XY) model is
\begin{equation}
H\{s\}= -\sum_x \sum_{ \mu=1{,}3 } s(x) \cdot s(x+e_\mu).
\label{eq:hamilt} \end{equation}
Here $s(x)$ is a two-component classical spin
of unit length associated
to the site  with position vector
$ x=n_1 e_1+n_2 e_2 +n_3 e_3 =(n_1,n_2,n_3)$ of a 3-dimensional
 simple cubic lattice
 and $ e_1$, $ e_2$, $ e_3$ are the  elementary lattice vectors.
The sum over $ x $
extends to all lattice sites.

It has been rigorously proved that the model exhibits a ferromagnetic
 phase transition [\cite{froeh}].

We present here series which extend by three terms,
 to order  $\beta^{17}$,
 the series of Ref.[\cite{luscher}].
 They  have been computed by a FORTRAN code
which iteratively solves the Schwinger-Dyson
equations for the correlation
 functions[\cite{bcm}].

We have tabulated
the HTE coefficients of the
two-point correlation function
\begin{equation}
 C(x;\beta)=<s(0)\cdot s(x)>
\label{eq:corr} \end{equation}
for all inequivalent sites $x$
for which the expansion is
 non trivial to order $\beta^{17}$.

We have analyzed the series for the spherical moments of
the correlation function  $m^{(l)}(\beta)$
 defined as follows:
\begin{equation}
 m^{(l)}(\beta)=\sum_x \vert x\vert^{l} C(x;\beta)
= \sum_{r=1}^\infty
 a^{(l)}_r \beta^r ,
\label{eq:moment} \end{equation}
(here $\vert x \vert =\sqrt{n_{1}^{2}+n_{2}^{2}+n_{3}^{2}}$ ),
$l \geq 0 $ and the sum extends over all lattice sites.
The zeroth order spherical moment  $m^{(0)}(\beta)$ is
 the (reduced)
susceptibility and is also denoted by $\chi(\beta)$.

In Table I, we report the HTE
coefficients of the spin-spin correlation function
$<s(0)\cdot s(x)>$ with
$x=(1,0,0) $.

In Tables II, III and IV we report the
expansion coefficients for the moments
$ m^{(l)}(\beta) $
with $l= 0,1$ and $2$.

Our analysis of this O(2) symmetric model parallels that of the
corresponding series for the O(0) symmetric
self-avoiding walk
(s.a.w.) model and the O(1) symmetric Ising model on the s.c.
lattice  [\cite{gutt}].  Using
both first-order
and second-order differential approximants, we first analyse the
susceptibility series.  We
find that if the degree of the inhomogeneous
polynomial is too low $(\leq 3)$ the
approximants cannot adequately accommodate the analytic background term.
On the other hand
if the degree of the inhomogeneous polynomial is too large $(\geq 8)$,
there are insufficient
series terms to adequately represent
the singular part of the series.  For intermediate
values of the degree of the inhomogeneous polynomial however, the
approximants are stable,
allowing the unbiased estimates $\beta_{c} = 0.45406 \pm 0.00005$, $\gamma
= 1.315 \pm 0.009$
to be made.  The unbiased estimates from second-order approximants were
more erratic, giving
$\beta_{c} = 0.4541 \pm 0.0001$, $\gamma = 1.32 \pm 0.01$.  Biasing the
approximants at
$\beta_{c} = 0.45406$ gave $\gamma = 1.315 \pm 0.005$ and $\gamma = 1.316
\pm 0.005$ from
first-order and second-order approximants respectively.
 The results remain
essentially unchanged using either the Fisher-Au Yang/Hunter-Baker
definition (no regular
singular point at the origin) or the  Guttmann/Joyce definition of the DAs
(a regular
singular point at the origin), the difference being mostly in the
dispersion of the data
(in particular of the background), which is somewhat greater
 with the former definition.

A similar analysis of the first moment, $m^{(1)}(\beta)$ is less
satisfactory.  Most of the
approximants are defective, but the few that are not are centred around a
slightly lower
temperature, $\beta_{c} = 0.4542 \pm 0.0003$, with exponent $\gamma + \nu =
2.00 \pm 0.03$.
Biasing the approximants at $\beta_{c} = 0.45406$ gives mainly defective
approximants,
slowly decreasing in value,
so that we can only estimate $\gamma + \nu \leq 2.00$.  The
second correlation moment series $m^{(2)}(\beta)$ is somewhat better
behaved, though, like
the analogous s.a.w. and Ising series,
unbiased approximants at first glance give a lower
critical temperature than do the susceptibility series approximants,
notably $\beta_{c} =
0.4542 \pm 0.0002$, and $\gamma + 2 \nu = 2.69 \pm 0.02$.  This behaviour
of the series
$m^{(2)}(\beta)$ was also noted for the Ising and s.a.w. model series
 [\cite{gutt}].  It appears
that longer series are needed for higher moments of the correlation
function.  Biasing the
second-moment series at $\beta_{c} = 0.45406$ gives $\gamma + 2 \nu =
2.67$, but this must
be regarded as an upper bound as the sequence of estimates of $\gamma + 2
\nu$ decreases
with increasing numbers of terms - just as observed previously for the
corresponding Ising
series.  While it is difficult to extrapolate this slowly declining
sequence, the limit
$2.66^{+0.01}_{-0.02}$ is likely sufficiently conservative to include the
correct value.  In
reaching this conclusion we have not only extrapolated this sequence, but
have studied the
behaviour of analogous sequences for the Ising and s.a.w. model, where we
also have exact
results for the two-dimensional models to guide us.  If this estimate is
accepted, we find
from our earlier estimate of $\gamma$ that $\nu = 0.67 \pm 0.01$.

We may also construct the series with
coefficients $c_r = m_r^{(2)}/m_r^{(0)}$
and study its singularity at $z=1$
which should have exponent $2\nu +1$.
Then we get  $\nu = 0.68 \pm 0.01$, with again a decreasing sequence of
exponent estimates,
suggesting that $\nu$ is in fact a little lower.

In conclusion our estimates of $\gamma$ are fairly precise
and, as shown later, in good
agreement with the Renormalization Group (RG)
 results.  However our estimates of $\nu$
cannot yet compete either with the precision of the experimental
data  nor with the RG  or Monte Carlo determinations, although they are
 perfectly compatible with both. This is probably due
to the slow convergence of $ m^{(2)}(\beta) $, and as noted above,
 has already been observed in the study
 of high order expansions for the SAW and Ising models [\cite{gutt}].
 Longer series are then required in order to make a
more accurate analysis possible  and in particulare to
 account properly for the confluent singularities.

Let us now briefly review previous high temperature
series analyses, restricting our review to the sc lattice results.

Bowers and Joyce [\cite{bowers}],
 computed series
 to order $\beta^8$ and gave the
following estimates:
$\beta_c=0.4530 \pm 0.0016 $, and $\gamma=1.312 \pm 0.006 $.

In Ref.[\cite{ferer}] the  series were extended
to order $\beta^{11}$. The estimated inverse critical
 temperature was $  \beta_c =0.4539 \pm 0.0013$ and the
 corresponding estimates for
 $ \gamma $ and $ \nu $ were
$ \gamma=1.32 \pm 0.05$ and  $ \nu=0.675 \pm 0.015$.
A comparison with our results shows that
our  central
 values for  $  \beta_c $ and $\gamma$
are  significantly lower and that
the precision in
our estimates has improved by a factor two.

Reliable Monte Carlo simulations with
good statistical accuracy,
on reasonably sized lattices,  have become possible only
recently after the invention of algorithms with
reduced critical slowing down [\cite{sokal}].
The largest accurately studied lattice is still
only $64^3$ sites large (present practical limits
seem to be around $100^3$ sites),
 which means that a very accurate treatment
of finite size effects is required and that the estimate of
systematic errors is very delicate.
The oldest analysis is due to  Li and Teitel [\cite{li}]
 who  performed a Metropolis simulation
( supplemented by over-relaxation
method ) on lattices up to $16^3$ sites.
A finite size scaling analysis of their data yields
$ \beta_c= 0.4533 \pm 0.0006$ and
$ \nu=0.67 \pm 0.02$.
(The model actually simulated is a clock model with 512 states.)

More recently Hasenbusch and Meyer [\cite{hasen}]
used the  Wolff single cluster algorithm
on lattices up to $96^3$ sites.  From a fit of the data to
$\chi \propto (\beta_c-\beta)^{-\gamma}$,
 they found $\beta_c=0.45421 \pm 0.00008$ and $\gamma= 1.327 \pm 0.008$.
A recent update [\cite{gott}] of this study  using the
  Wolff single cluster algorithm
on lattices up to $64^3$ sites gave
$\beta_c=0.45420 \pm 0.00002,
 \nu = 0.664 \pm 0.006$ and $\gamma= 1.324 \pm 0.001$.

W. Janke [\cite{janke}]
 also used the Wolff single cluster algorithm
on lattices up to $48^3$ sites.  From a study of the fourth order cumulant
he obtained
 $\beta_c=0.4542 \pm 0.0001$ and  $\nu= 0.670 \pm 0.002$.
Fitting data to the formula $\chi \propto \chi_+ (\beta_c-\beta)^{-\gamma}$
he obtained
$\beta_c=0.45408 \pm 0.00008$,
 and $\gamma= 1.316 \pm 0.005$.
 Repeating his fit with  fixed $\beta_c= 0.4542$ the value of
  $ \gamma$ increases to $ \gamma= 1.323 \pm 0.002$.

The previous computations should also be compared to the estimates
by the Renormalization Group applied to
an O(2) symmetric $\phi ^4$ field theory model.

Sixth order perturbation expansion in three dimensions
by Baker, Nickel and Meiron  [\cite{baker}],
gave
$\gamma = 1.316 \pm 0.009$ and
$\nu = 0.669 \pm 0.003$.
Subsequently, taking into account the large order
behavior of the perturbation
series coefficients,  Le Guillou  and
Zinn Justin  [\cite{zinn}] refined these estimates and obtained
 $ \gamma = 1.316 \pm 0.0025$ and $\nu = 0.6695 \pm 0.001 $.

Performing the computation[\cite{guillou}] by the
Wilson-Fisher $\epsilon = 4-d$
 expansion  Borel resummed to order $\epsilon^5$, Le Guillou and Zinn
Justin subsequently
obtained the following estimates:
$\gamma = 1.315 \pm 0.007$ and $ \nu = 0.671 \pm 0.005$ .

It thus appears that the RG results for
$\gamma $ are slightly smaller
 than the old HT and some of the new
Monte Carlo estimates, but  perfectly compatible with the results of our
 analysis, while our estimate of $\nu$ is compatible with, but less
accurate than, the most
recent RG results.

\nonum
\section{Acknowledgments}
\label{ack}
Our work has been partially supported by MURST.  AJG wishes to acknowledge
support from the
Australian Research Council.

\narrowtext
\begin{table}
\caption{HTE coefficients of the
nearest neighbor  correlation  $C(0,x)$ with $x=(1,0,0)$}\label{Cnn}
\begin{tabular}{cc}
order&coefficient\\
\tableline
          1  &0.50000000000000000000000000000000\\
          3  &0.43750000000000000000000000000000\\
          5  &1.01041666666666666666666666666667\\
          7  &2.49169921875000000000000000000000\\
          9  &7.48240559895833333333333333333333\\
         11  &24.7292479338469328703703703703704\\
         13  &86.7042412409706721230158730158730\\
         15  &317.800753506891941898083560681217\\
         17  &1205.06602454131493586174488161069\\
\end{tabular}
\end{table}

\narrowtext
\begin{table}
\caption{HTE coefficients of the
susceptibility $m^{(0)}$.}\label{mom0}
\begin{tabular}{cc}
order&coefficient\\
\tableline
	0    &1.00000000000000000000000000000000\\
	1    &3.00000000000000000000000000000000\\
	2    &7.50000000000000000000000000000000\\
	3    &18.3750000000000000000000000000000\\
	4    &43.5000000000000000000000000000000\\
	5    &102.343750000000000000000000000000\\
	6    &237.054687500000000000000000000000\\
	7    &546.946289062500000000000000000000\\
	8    &1252.00488281250000000000000000000\\
	9    &2858.81752929687500000000000000000\\
	10   &6496.15140787760416666666666666666\\
	11   &14735.3746412489149305555555555555\\
	12   &33314.7537746853298611111111111111\\
	13   &75222.2566392081124441964285714286\\
	14   &169444.488235923222133091517857143\\
        15   &381306.311343971793613736591641865\\
        16   &856543.263379992410619422872230489\\
        17   &1922537.91945074856684251367029620\\

\end{tabular}
\end{table}
\narrowtext
\begin{table}
\caption{HTE coefficients of the
first correlation moment $m^{(1)}.$}\label{mom1}
\begin{tabular}{cc}
order&coefficient\\
\tableline
0    &0.00000000000000000000000000000000\\
1    &3.00000000000000000000000000000000\\
2    &11.4852813742385702928101323452582\\
3    &35.3919166429113710288472410676167\\
4    &100.391645797382835211177404733391\\
5    &270.169140885332810622174619548742\\
6    &703.928165009702962171567107355945\\
7    &1789.19653133764917963889959830865\\
8    &4468.32789180460469625866305929854\\
9    &11000.8726669685811734428842857616\\
10   &26788.0560947846126416923232831814\\
11   &64627.3429637161982839763298977200\\
12   &154749.818273239775925845196634614\\
13   &368132.797893714045088109726930470\\
14   &870977.871997489895140365839695762\\
15   &2050710.75491296809029207572988208\\
16   &4808405.28831745065018387682551317\\
17   &11232374.4966970585972846903939187\\

\end{tabular}
\end{table}
\narrowtext
\begin{table}
\caption{HTE coefficients of the
second correlation moment $m^{(2)}$.}\label{mom2}
\begin{tabular}{cc}
order&coefficient\\
\tableline
0	&0.00000000000000000000000000000000\\
1	&3.00000000000000000000000000000000\\
2	&18.0000000000000000000000000000000\\
3	&72.3750000000000000000000000000000\\
4	&247.500000000000000000000000000000\\
5    &770.593750000000000000000000000000\\
6    &2261.34375000000000000000000000000\\
7    &6360.66503906250000000000000000000\\
8    &17343.7773437500000000000000000000\\
9    &46158.4210449218750000000000000000\\
10   &120515.319303385416666666666666667\\
11   &309746.425031873914930555555555556\\
12   &785831.296427408854166666666666667\\
13   &1971809.99205790928431919642857143\\
14   &4901417.59164962163047185019841270\\
15   &12084656.3170853394364553784567212\\
16   &29584235.7640201335230832377438823\\
17   &71970593.8709586784015817546900548\\
\end{tabular}
\end{table}
\end{document}